# The Network-based Candidate Forwarding Set Optimization Approach for Opportunistic Routing in Wireless Sensor Network

Ning Li, *Member IEEE*, Alex X. Liu, *Fellow IEEE*, Jose-Fernan Martinez-Ortega, Vicente Hernandez Diaz, Xin Yuan

*Abstract*-In wireless sensor networks (WSNs), the opportunistic routing has better performances on packet delivery probability than the deterministic routing. For reducing the transmission delay and duplicate transmission in opportunistic routing, the candidate forwarding set optimization technology is proposed. This technology is crucial to opportunistic routing. The prior arts have limitations on improving the routing performances. Therefore, in this paper, we propose the concept of the network-based approach to address the disadvantages of prior arts. In the network-based approach, the nodes in the candidate forwarding set are divided into different fully connected relay networks. For the forwarding nodes in CFS, more than one relay network can be constructed and only one relay network can be chosen as the final relay network. So, first, we propose the fully connected relay network judgment algorithm to judge whether the chosen network is fully connected or not. Next, the properties of these networks are investigated. In this algorithm, the relay network selection takes the packet delivery probability between the sender and the relay networks, the transmission delay, and the forwarding priorities of nodes in relay networks into account to choose the most effective relay network. The nodes in this relay network will be chosen as final forwarding nodes. By these, the transmission delay and duplicate transmission are reduced while the efficiency of opportunistic routing is kept. Based on these innovations, the proposed algorithm can improve network performances greater than that of ExOR and SOAR.

*Index Term*-Opportunistic routing, wireless sensor networks, candidate forwarding set, transmission delay, duplicate transmission.

## I. INTRODUCTION

### A. Motivation

In wireless sensor networks (WSNs), compared with the deterministic routing, the opportunistic routing has better performance on packet delivery probability. In opportunistic routing, the packet delivery probability is defined as the probability that the data packet sent by the sender can be received successfully by at least one node in the candidate forwarding set (CFS) [1][2]. This is important to the wireless sensor network because the routing algorithms guarantee effective and reliable data transmission from the source node to the destination node. However, since more than one neighbor can receive the data packet from the sender in opportunistic routing, the transmission delay and duplicate transmission are more serious than that in deterministic routing [3][4][5]. For improving the performance of opportunistic routing, the candidate forwarding set optimization technology is proposed. The candidate forwarding set optimization means to remove some forwarding nodes from the CFS based on the requirements of routing performance [5][6][7].

The candidate forwarding set optimization is crucial to the opportunistic routing. Because the candidate forwarding set optimization has a great effect on the transmission delay, the duplicate transmission, and the packet delivery probability. The CFS optimization includes three aspects. First, the number of forwarding nodes in the CFS should be optimized [5]. In opportunistic routing, the more forwarding nodes in CFS, the higher packet delivery probability. However, if there are too many nodes in the CFS, the overhead, the duplicate transmission, and the transmission delay become serious. Second, the forwarding nodes in the CFS should be fully connected to reduce the transmission delay and duplicate transmission [8][9]. Third, the higher-priority forwarding nodes should be selected as much as possible to improve the routing performance (this will be proved in the following of this paper). Three different kinds of CFS optimization approaches have been proposed in the past decades. The most commonly used approach is to remove the candidates that are worse than the sender according to a specific metric [8][9], such as packet delivery probability, duplicate likeliness, and node contribution; the second approach consists in filtering out the bad candidates based on their duplicate forwarding probability [10]; the third method removes the candidate relays with low contribution in traffic forwarding [11-13]. All these algorithms are called *node-based approach* because they optimize the CFS based on the parameters of single forwarding nodes. So, the properties of the whole CFS cannot be considered during the routing process and the overall routing performance cannot be guaranteed. This is also the main disadvantage of the node-based approach. Therefore, in this paper, we propose the concept of the *network-based approach* to address the disadvantages of node-based approach. To the best of our knowledge, the network-based approach has not been investigated sufficiently in the previous works.

### B. Problem Statement

Ning Li and Alex X. Liu are with the Dongguan University of Technology, Dongguan, China. E-mail: {li.ning@upm.es, alexliu@dgut.edu.cn}.
Ning Li, Jose-Fernan Martinez-Ortega, and Vicente Hernandez Diaz are with the Universidad Politecnica de Madrid, Madrid, Spain. E-mail: {li.ning, jf.martinez, vicente.hernandez}@upm.es.
Xin Yuan is with the Harbin Institute of Technology, Weihai, China. E-mail: xin.yuan@hit.edu.cn.

The research leading to the presented results has been undertaken with in the SWARMs European project (Smart and Networking Underwater Robots in Cooperation Meshes), under Grant Agreement n. 662107-SWARMs-ECSEL-2014-1, partially supported by the ECSEL JU and the Spanish Ministry of Economy and Competitiveness (Ref: PCIN-2014-022-C02-02).

So, the main objective of this paper is to propose the network-based CFS optimization approach for opportunistic routing in wireless sensor networks (WSNs). The network-based CFS optimization approach includes two fundamental problems. The first problem is how we construct and judge the relay networks based on the nodes in CFS. The second problem is how we choose the most appropriate relay network to improve the performances of opportunistic routing, such as reducing the transmission delay, energy consumption, and duplicate transmission of network, etc. In this paper, the network-based CFS optimization approach is proposed for time-based coordination scheme; because the performances of time-based coordination scheme (including the transmission delay and duplicate transmission) are better than the other coordination schemes [22-25].

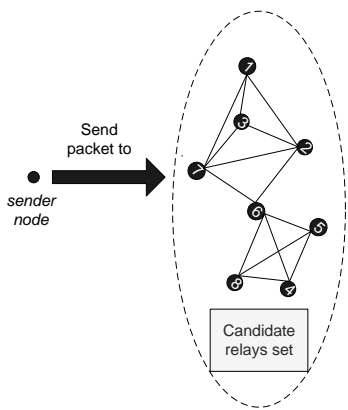

Fig. 1. The candidate relay networks of opportunistic routing.

Specifically, the problems will be addressed in this paper are shown as follows. The *first* problem needs to be addressed is to judge whether the relay network constructed by candidate forwarding nodes is fully connected or not. As shown in Fig. 1, there are many forwarding nodes in CFS, so many fully connected relay networks can be constructed, such as network (1,2,3,7), network (4,5,8), etc. Therefore, for the given candidate forwarding nodes, whether the networks constructed by these nodes are fully connected or not need to be investigated. The *second* problem needs to be addressed is to choose the most appropriate relay network from all the fully connected relay networks. For the CFS shown in Fig. 1, many different fully connected relay networks can be constructed. The nodes and topologies in these relay networks are different, such as network (1,2,3,7) and network (4,5,8), etc. So, the properties (i.e., the relaying delay, the packet delivery probability, etc.) of these networks are also different. For example, the packet delivery probability and the relaying delay of network (1,2,3,7) and network (4,5,8) are different. Therefore, to evaluate the performance of these relay networks and select the most appropriate one for opportunistic routing are also the targets of this paper. The *third* problem needs to be addressed is to investigate the effect of forwarding nodes' packet delivery probabilities on routing performance. Because packet delivery probabilities of different nodes in communication link have different effects on routing performance [1][14]. So, in this paper, we will investigate how the routing performances, such as the transmission delay, are affected by the forwarding nodes' packet delivery probabilities in the communication link. The *final* problem needs to be addressed is to combine CFS optimization with the nodes' forwarding priority in CFS. Because the nodes' forwarding priority has a great effect on the routing performance. For example, CFS optimization based on the connectivity criterion removes candidate forwarding nodes that trigger duplicate transmissions. However, these removed nodes may be the best in terms of the routing metric (i.e., has the highest forwarding priority) and may bring the highest expected performance. Therefore, the routing performance will be affected. So, it is necessary to jointly consider the node's forwarding priority and CFS optimization.

Overall, in our network-based CFS optimization approach, the following parameters or performance metrics are considered. The first parameter is the full connectivity of the relay network, i.e., the relay network should be fully connected. The second parameter is the forwarding priority of the forwarding nodes. The forwarding priority is an important parameter in time-based coordination scheme. In opportunistic routing, the forwarding priority is calculated based on one or more parameters of the forwarding node. For instance, the node whose packet delivery probability is high will have high forwarding priority. So, the forwarding priority can reflect the forwarding node's properties. The third parameter is the transmission delay and duplicate transmission of the relay network. Note that this parameter is the transmission delay and duplicate transmission of the relay network rather than that of the forwarding node. The fourth parameter is the packet delivery probability of the relay network. Note that this parameter is the packet delivery probability of the relay network rather than the forwarding node.

*C. Limitations of Prior Arts*

Because the node-based approach optimizes CFS based on the parameters of single forwarding nodes but the routing performance relates to all the forwarding nodes in CFS, so the node-based approaches have limited capability on improving the overall routing performance. There are three limitations that are difficult addressed based on the node-based CFS optimization approach. *First*, for reducing the transmission delay and duplicate transmission of the time-based coordination scheme, the forwarding nodes in CFS should be fully connected. However, the previous CFS optimization algorithms do not take this requirement into account. Even some previous works, such as [8] and [9], take the connectivity into account, they have the limitations introduced following. *Second*, the node-based CFS optimization approach cannot reflect the overall characteristics of the CFS. For instance, if the node-based approach takes the full connectivity into account, based on this approach, only one fully connected relay network can be found at each time; assume this relay network is network (1,2,3,7). This approach cannot find all the fully connected relay networks constructed by the nodes in CFS. However, as introduced in Section I.B, many fully connected relay networks can be constructed, and the other relay networks may be better than network (1,2,3,7). The node-based approach cannot address this problem. *Third*, in opportunistic routing, the forwarding priorities of different forwarding nodes have a great effect on routing performance. However, the CFS optimization and the calculation of the forwarding nodes' forwarding priority are done separately in node-based approach.

Except for the limitations introduced above, there are also some properties which are important to the routing performance have not been investigated in the node-based CFS optimization approach. The *first* is that the effect of forwarding nodes' priorities on routing performance. As the viewpoints in [1] and [14], different forwarding nodes in the different positions of communication link have different effects on routing performances. For instance, the packet delivery probability of forwarding node at the end of the communication link has a great effect on energy consumption and transmission delay [14]; the ETX relates to all the packet delivery probabilities of nodes in communication link [1]. However, in previous works, this effect on routing performance has not been investigated. The *second* is that how to judge whether the network constructed by the nodes in CFS is fully connected or not. The *third* is how to choose the most appropriate relay network. As shown in Fig. 1, for the nodes in CFS, many fully connected relay networks can be constructed. The nodes and the topologies of these networks are all different. This means that the properties of these networks are also different. Different relay networks have different effects on routing performances. However, how to choose the most appropriate one from these networks has not been investigated in previous works.

### D. Proposed approach and advantages over prior arts

Based on the problems and the limitations presented in Section I.B and Section I.C, we propose the network-based delay and duplicate transmission avoid (DDA) CFS optimization algorithm for time-based coordination scheme. This algorithm is proposed for wireless sensor networks. In DDA, the nodes in the candidate forwarding set are divided into different fully connected relay networks. For the forwarding nodes in CFS, more than one relay network can be constructed and only one relay network can be chosen as the final relay network. So, first, we propose the fully connected relay network recognition algorithm to judge whether the chosen network is fully connected or not. Next, the properties of these networks are investigated. In DDA, the relay network selection takes the packet delivery probability between the sender and relay networks, the transmission delay, and the forwarding priorities of nodes in relay networks into account to choose the most effective relay network. The nodes in this relay network will be chosen as final forwarding nodes. By these, the transmission delay and duplicate transmission are reduced while the efficiency of opportunistic routing is kept.

Our proposed network-based CFS optimization algorithm can overcome the limitations of the node-based approach. First, the proposed network-based approach takes the full connectivity of relay networks into account to reduce the duplicate transmission of time-based coordination scheme. Therefore, the performances of the network-based approach are better than that of the node-based approach. For instance, the duplicate transmission in DDA is reduced greatly in the network-based approach compared with the node-based approach. Second, the proposed network-based approach combines the forwarding priorities of forwarding nodes with CFS optimization. Based on this, the forwarding node's properties are taken into account during CFS optimization. Moreover, compared with the node-based approach, the effect of forwarding nodes' priorities on routing performance, how to judge whether the network constructed by candidate forwarding nodes is fully connected or not, and how to choose the most appropriate relay network are investigated in the proposed network-based CFS optimization approach.

### E. Technical Challenges and Solutions

These are some technical challenges to address the problems and limitations of prior arts. The main technical challenge is to propose the concept of network-based CFS optimization approach. For perfection the theory of network-based approach, three technical challenges appear. The *first technical challenge* is to judge whether any *n* nodes in CFS can construct a fully connected relay network or not. This is a technical challenge because: on one hand, for the forwarding nodes in CFS, many fully connected relay networks can be constructed, so it is difficult to judge whether the give *n* forwarding nodes can construct a fully connected network or not; on the other hand, the proposed fully connected network judgment algorithm should effective and simple to reduce the routing overhead as much as possible. For addressing this challenge, in this paper, we propose a neighbor matrix based fully connected network recognition algorithm, which is effective and simple. The *second* technical challenge is to explore the properties of the fully connected relay network. This is technical challenge because: on one hand, the forwarding nodes in CFS can construct many fully connected relay networks; the nodes and topologies of these relay networks are all different; on the other hand, many parameters of forwarding nodes (such as the packet delivery probability, the forwarding priority, etc.) can affect the properties of relay network. To address this challenge, in this paper, we investigate the properties of the relay network in detail. The *third* technical challenge is to choose the most appropriate relay network. This is a technical challenge because for the forwarding nodes in CFS, many fully connected relay networks can be constructed; the properties of these networks are all different, so it is challenging to choose the best one from these networks to improve the routing performance. To address this challenge, in this paper, we introduce the multi-attribute utility theory into the chosen of the best relay network. In this algorithm, not only the transmission delay and packet delivery probability but also the node utility and the forwarding priority of node are considered. The *final* technical challenge is to combine the CFS optimization with the forwarding node utility which is used to determine the forwarding priority. This is a technical challenge because these two parameters are calculated in different stages of opportunistic routing. So, if these two parameters are optimized jointly, the extra overhead will be needed. This will increase the complexity of the algorithm. However, our proposed network-based CFS optimization approach can optimize these two parameters jointly while the overhead and complexity of opportunistic routing do not increase.

## II. RELATED WORKS

There are some CFS optimization algorithms have been proposed in the past decades. These CFS optimization algorithms are all node-based approach. The works in [8] and [9] optimize the candidate forwarding nodes in CFS based on the connectivity. The nodes are connected in a mesh fashion will be selected as the relay nodes. However, only one fully connected relay network can be constructed in these two algorithms at each time. Except for the connectivity, some works filter out the bad candidates based on their duplicate forwarding probability. This probability was derived using a discrete-time Markov model in [10]. The candidate forwarding nodes which have high duplicate forwarding probability will be deleted. However, in this approach, the transmission delay, the connectivity of forwarding nodes, and the forwarding priority of forwarding nodes are not considered. The third candidate filter method is to remove the candidates with a low contribution in data transmission [11][12][13]. For example, in MORE [11] and CCACK [12], each forwarding node is given a predicted number of transmissions for a flow. If a candidate does not perform at least 10% of the overall predicted transmissions, it is deleted. The main disadvantage of these works is that they fail to take the forwarding priority and the connectivity of forwarding nodes into account when optimizing the candidate forwarding nodes.

Another candidate optimal approach of CFS that adopted in some previous works is the optimization approach which can determine the optimal CFS directly without applying the candidate filter technique [15][16][17]. For instance, in [15], the authors propose an optimal algorithm that can generate the candidate relay sets by a Dijkstra-like algorithm. In LCAR [16], the authors propose a shortest any-path algorithm to find the optimal candidate relay set; in this algorithm, finding the shortest any-path is based on a generalization of the Bellman–Ford algorithm. However, these algorithms fail to address the problem of redundant transmissions caused by node overhearing; thus, the redundant transmission and the transmission delay are serious in these algorithms.

Moreover, since the above CFS optimization algorithms are all node-based approach, so they all have the limitations introduced in Section I.C.

### III. NETWORK MODEL

*A. Network model*

In the network model used in this paper, the nodes cannot move or can move slowly[1]. Two nodes can communicate with each other directly (without the help of the third node) if and only if there is a bi-directional communication link between these two nodes. The bi-directional communication link means that the transmission ranges of these two nodes are all larger than the distance between these two nodes. For instance, as shown in Fig. 2(a), node *s* and node 7 can communicate with each other directly when $\|s7\| \leq r_s$ and $\|s7\| \leq r_7$, where $\|s7\|$ is the Euclidean distance between node *s* and node 7, $r_s$ and $r_7$ are the transmission ranges of node *s* and node 7, respectively. The transmission range of node *s* is a circle which the center is node *s* and the radius is $r_s$, denoted as $C(s, r_s)$. This can be found in Fig. 2(a).

As shown in Fig. 2(a), in opportunistic routing, when the sender wants to send a data packet, first, a set of neighbor nodes are chosen as candidate forwarding nodes based on some performance metrics (such as ETX, distance, etc.). Then, the sender sends the data packet to all the nodes in candidate forwarding set $\mathbb{R}$ (the candidate forwarding set is the set of all the candidate forwarding nodes). For instance, in Fig. 2(a), $\mathbb{R} = \{1,2,3,4,5,6,7,8\}$. The network constructed by the nodes in $\mathbb{R}$ is denoted as $G(V_\mathbb{R}, E_\mathbb{R})$, where $V_\mathbb{R}$ represents the set of nodes in $\mathbb{R}$ and $E_\mathbb{R}$ represents the set of bi-directional communication links in the network. Second, the candidate forwarding nodes relay the data packet to the next hop candidate forwarding nodes with the same process as the sender.

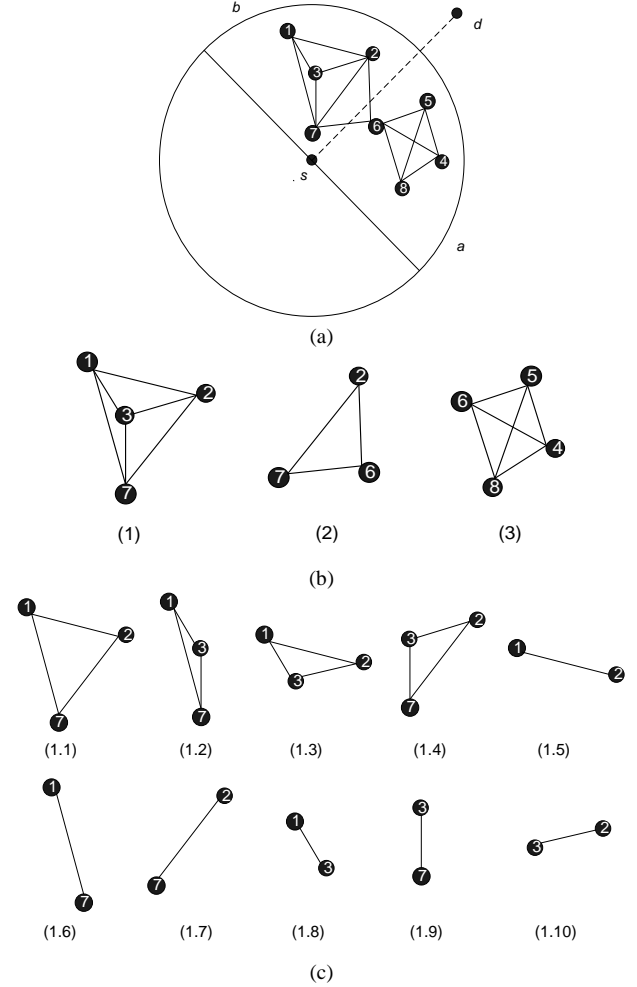

Fig. 2. The network model for opportunistic routing: (a) the network of the candidate forwarding nodes; (b) the independent sub-networks of the original network; (c) the dependent sub-networks of Fig. 2(b.1)

In the first step, the candidate forwarding set needs to be optimized. For instance, in time-based coordination scheme, for reducing transmission delay and duplicate transmission, the forwarding nodes should be able to communicate with each other directly, i.e., the network constructed by these nodes should be fully connected. The fully connected network means that there exists a bi-directional communication link between any two nodes in this network; otherwise, the network is not fully connected. However, as shown in Fig. 2(a), the $G(V_\mathbb{R}, E_\mathbb{R})$ may not the fully connected network. For instance, node_3 and node_6 cannot connect directly. One feasible approach is to keep the fully connected candidate forwarding node set $\mathbb{R}^*$ and remove the un-fully connected nodes, where $\mathbb{R}^*$ is the subset of $\mathbb{R}$. For instance, the nodes in $\mathbb{R}^* = \{1,2,3,7\}$, which is shown in Fig. 2(b), are fully connected. To the candidate forwarding set $\mathbb{R}$, there are many different subsets $\mathbb{R}^*$. This means that to $G(V_\mathbb{R}, E_\mathbb{R})$, there are many fully connected sub-networks $G(V_{\mathbb{R}^*}, E_{\mathbb{R}^*})$ can be constructed by the nodes in $\mathbb{R}$. For example, the networks are shown in Fig. 2(b) and Fig. 2(c) are all the fully connected sub-networks of Fig. 2(a). Since these fully connected networks are different, for investigating the differences between these networks more clearly, the following definitions are presented.

---

[1]Because the time needed for data transmission is extremely short, so when the node moves slowly, the node movement has a little effect on routing performance. However, when the nodes move quickly, our proposed algorithm will deteriorate.

In the fully connected networks, there must have bi-directional links between any two nodes. So, we can simplify the expression of the fully connected network by only showing the nodes in this network. For instance, for $G((2,6,7),(\overrightarrow{26},\overrightarrow{27},\overrightarrow{67}))$ that shown in Fig. 2(b.2), we can simplify the expression as $G(2,6,7)$. Since the relay networks that constituted by candidate forwarding nodes should be fully connected to reduce the transmission delay and duplicate transmission, we define the relay network as follows.

**Definition 1:** For candidate forwarding set $\mathbb{R}$, the relay networks are defined as the fully connected sub-networks of $G(V_\mathbb{R})$, denoted as $G(V_{\mathbb{R}^*})$.

For instance, in Fig. 2(a), $G(2,6,7)$ is one of the relay networks. Since there are more than one relay network and the nodes in these relay networks are different, such as the relay networks $G(2,6,7)$ and $G(1,2,3,7)$, for distinguishing these networks, we define the network degree in Definition 2.

**Definition 2:** The degree of the relay network is defined as the number of nodes in this relay network, denoted as $d_G$.

For instance, in Fig. 2(b), the network degree of Fig. 2(b.1) is 4. Note the fact that in relay networks, the small degree relay networks may be the sub-network of a large degree relay networks (it is not always true). So, we define the relevant and irrelevant for the relay networks in Definition 3.

**Definition 3:** For any two relay networks $G(V_{\mathbb{R}_1^*})$ and $G(V_{\mathbb{R}_2^*})$, in which $V_{\mathbb{R}_1^*} \notin V_{\mathbb{R}_2^*}$ and $V_{\mathbb{R}_2^*} \notin V_{\mathbb{R}_1^*}$, if $G(V_{\mathbb{R}_1^*} + V_{\mathbb{R}_2^*})$ is still a relay network, then these two relay networks are relevant; otherwise, these two relay networks are irrelevant.

For instance, for relay networks $G(1,2,3)$ and $G(2,3,7)$, since $G(1,2,3,7)$ is still the relay network, $G(1,2,3)$ and $G(2,3,7)$ are relevant; for relay networks $G(2,6,7)$ and $G(1,2,3,7)$, since $G(1,2,3,6,7)$ is not the relay network, $G(2,6,7)$ and $G(1,2,3,7)$ are irrelevant. Note that $G(1,2,3,7)$ and $G(1,2,3)$ are not relevant, since $(1,2,3) \in (1,2,3,7)$. Based on Definition 3, we can give Definition 4 as follows.

**Definition 4:** For relay network $G(V_{\mathbb{R}_i^*})$, if there exists a relay network $G(V_{\mathbb{R}_j^*})$ which is relevant with $G(V_{\mathbb{R}_i^*})$, then $G(V_{\mathbb{R}_i^*})$ is called *s-network*; otherwise, $G(V_{\mathbb{R}_j^*})$ is called *o-network*.

For instance, the $G(1,2,3,7)$ shown in Fig. 2(b) is an *o*-network; the $G(1,2,3)$ shown in Fig. 2(c) is a *s*-network of $G(1,2,3,7)$. The *s*-network can be derived from the *o*-network. For each *o*-network, there are more than one *s*-networks can be derived from this *o*-network. The degrees of these *s*-networks are smaller than that of the *o*-network. For instance, the relay networks shown in Fig. 2(c) are all *s*-networks that derived from the *o*-network shown in Fig. 2(b.1). Moreover, since the network degree of Fig. 2(b.1) is 4, so the *s*-networks that derived from Fig. 2(b.1) will be 2-degree and 3-degree, respectively. Note that if the network degree is 1-degree, then the algorithm will be the same as the deterministic routing, so in this paper, we do not consider the 1-degree networks. The notations used in this paper are listed in Table 1.

TABLE 1
THE NOTATIONS USED IN THIS PAPER

| parameter | meaning |
|---|---|
| $\mathbb{R}$ | the candidate forwarding set before optimizing |
| $\mathbb{R}^*$ | the final candidate forwarding set after optimizing |
| $V_{\mathbb{R}_1^*}$ | the set of nodes in $\mathbb{R}_1^*$. |
| $T$ | waiting time in time-based coordination scheme |
| $DT_{G(1,2,\ldots,n)}$ | relaying delay of the relay network $G(1,2,\ldots,n)$ |
| $P_{G(1,2,\ldots n)}$ | packet delivery probability of relay network $G(1,2,\ldots,n)$ |
| $P_i$ | the packet delivery probability of the *ith* priority node in $\mathbb{R}^*$ |
| $\Delta DT^i_{G(1,2,\ldots n)}$ | the variation of $DT_{G(1,2,\ldots,n)}$ when the packet delivery probability of *ith* priority node changes |
| $\Delta DT^{(i,j)}_{G(1,2,\ldots n)}$ | the difference of the $DT_{G(1,2,\ldots,n)}$ variation between any two forwarding nodes in $\mathbb{R}^*$ |
| $U$ | the node utility calculated when determining the forwarding priority of a node |
| $ETX_{one-hop}$ | one-hop ETX for each forwarding node in $\mathbb{R}^*$ |
| $U_i^*$ | the node utility of the forwarding nodes in $\mathbb{R}^*$ when taking the $ETX_{one-hop}$ into account |
| $neib_i$ | the neighbor matrix of *ith* node in $\mathbb{R}$ |
| $D_{G(1,2,\ldots,n)}$ | the result of (10) of the relay network $G(1,2,\ldots,n)$ |
| $t_{G(1,2,\ldots,n)}$ | the one-hop ETX of the relay network $G(1,2,\ldots,n)$ |
| $DT^*_{G(1,2,\ldots,n)}$ | network relaying delay when taking $t_{G(1,2,\ldots,n)}$ into account |
| $U_{G(1,2,\ldots,n)}$ | the utility of the relay network $G(1,2,\ldots,n)$ |
| $U^*_{G(1,2,\ldots,n)}$ | the utility of the relay network $G(1,2,\ldots,n)$ when taking $t_{G(1,2,\ldots,n)}$ into account |
| $U^F_{G(1,2,\ldots,n)}$ | the final utility of relay network $G(1,2,\ldots,n)$ |
| $v_{rx}$ | the relative variance of parameter $x$ |

*B. The relaying delay and packet delivery probability of the network*

For investigating the performance of the relay networks, in this section, we will introduce the calculation model of relaying delay and packet delivery probability of the relay network. There are two different kinds of packet delivery probabilities used in this paper: the packet delivery probability of the forwarding node and the packet delivery probability of the relay network. In the following, we will define these two probabilities.

**Definition 5:** The packet delivery probability of the forwarding node is defined as the probability that the data packet sent by the sender can be received successfully by node *i* in $\mathbb{R}^*$, denoted as $P_i$; the packet delivery probability of the relay network is defined as the probability that the data packet sent by the sender can be received successfully by at least one forwarding node in $\mathbb{R}^*$, denoted as $P_{G(1,2,\ldots n)}$.

According to Definition 5, the packet delivery probability of the relay network $G(1,2,\ldots n)$ can be calculated as [1]:
$$P_{G(1,2,\ldots n)} = 1 - \prod_{i=1}^n (1 - P_i). \qquad (1)$$

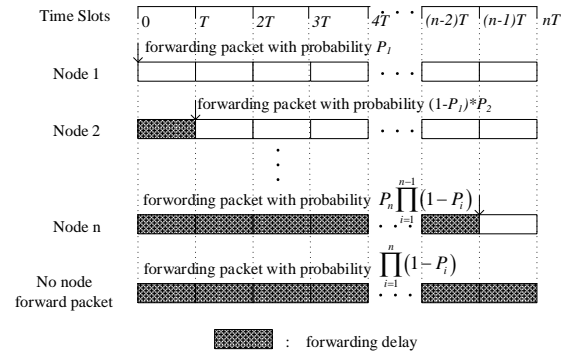

Fig. 3. The principle of the time-based coordination scheme

For the time-based coordination scheme, the relaying delay is mainly caused by overhearing the higher-priority node's ACK message [1]. For better understanding the relaying delay of the time-based coordination scheme, in the following, we introduce the principle of the time-based coordination scheme in detail. The principle can be found in Fig. 3.

As shown in Fig. 3, before data transmission in time-based coordination scheme, the forwarding nodes in CFS will be set forwarding priority based on the node utility $U$. The node utility is calculated based on one or more parameters of the forwarding nodes, such as the packet delivery probability, the transmission delay, etc. Then the sender will send the data packet to all the forwarding nodes in CFS. The higher-priority node has a higher priority to relay data packet to the next hop forwarding nodes than that of the lower-priority nodes; the lower-priority nodes overhear the ACK messages from the higher-priority nodes [1]. In CFS, the first-priority node[2] will check if it receives the data packet. If yes, this node will be the new sender immediately and broadcasts the ACK message to other candidate forwarding nodes; the candidate forwarding nodes which receive this message will drop the data packet that received from the sender. If the first-priority node fails to receive the data packet, then after time $T$ (which is called the waiting time, in [8], this time is set to 45*ms*), the second-priority forwarding node begins the same process as the first-priority node. This process will be repeated until one of the forwarding nodes receives the data packet or none of the nodes receives the data packet. Therefore, the average one-hop relaying delay after one transmission try can be calculated as:

$$DT_{G(1,2,\ldots n)} = \left(\sum_{i=1}^{n-1} iP_{i+1} \prod_{j=1}^{i}(1-P_i) + n\prod_{i=1}^{n}(1-P_i)\right)T \quad (2)$$

where $n$ is the degree of the relay network, $i$ is the priority of each node in the relay network, $P_i$ is the packet delivery probability of the *ith*-priority node in $\mathbb{R}^*$ and $0 < P_i < 1$, $T$ is the waiting period. The second term in (2) represents that none of the nodes receives the data packet that transmitted from the sender. Based on the average one-hop relaying delay introduced in (2), we have the corollary as follows.

**Corollary 1.** To the same relay network, the network relaying delay will be the smallest when the node priorities are determined based on the packet delivery probability of node.
**Proof.** See Appendix A.

From (1) and (2), we can conclude that even the *s*-networks can be derived from the *o*-networks, the relaying delay and the network packet delivery probabilities of these two kinds of networks are different. In the next section, we will investigate the properties of the relay networks in detail.

## IV. PROPERTIES OF THE RELAY NETWORK

In this section, based on the calculation model of network relaying delay and network packet delivery probability that proposed in Section III, we investigate the properties of the relay network in detail. The properties are divided into *in-network properties* and *inter-network properties*. These network properties can be used during the relay network selection. In the opportunistic routing, for determining the priorities of the candidate forwarding nodes, different performance metrics are used based on different application purposes. These metrics can be divided into two different categories: 1) the packet delivery probability based metrics, such as the ETX [1], the link correlation [8], etc.; and 2) not the packet delivery probability based metrics, such as the distance to the destination nodes, the residual energy, the interference, etc. The network relaying delay of these two different routing algorithms is different, since the network relaying delay is affected seriously by the packet delivery probability of the forwarding nodes and their forwarding priorities. This will be proved in the following of this section. As shown in Corollary 1, to the same network, the network relaying delay will be different when the node priorities are different. However, as shown in (1), to the same relay network, the packet delivery probability of this relay network is the same even the node priorities are changed.

### A. Inter-network properties

The inter-network properties represent the properties of the whole relaying network, i.e., the relaying network is regarded as an entirety.
**Corollary 2:** If the $G(V,E)$ is a relay network, then $E = V(V-1)/2$; otherwise, $E < V(V-1)/2$.
**Proof.** See Appendix B.

For each $\mathbb{R}$, in which the number of forwarding nodes is $n$, the number of the relay networks (including the *s*-networks and the *o*-networks) can be calculated as:

$$num = \sum_{i=2}^{n-1} c_n^i. \quad (3)$$

In (3), $c_n^i$ is the number of *i*-degree relay networks. In this paper, the 1-degree network has been ignored, since the 1-degree network is equal to the deterministic routing.

### B. In-network properties

In the relay networks, different node parameters, including the packet delivery probability and the node priority, have different effects on the network performance. For investigating the effect of the node parameters on network performances, in this section, we investigate the in-network properties of the relay network.

**Definition 6:** To the relay network $G(1,2,\ldots n)$, the effect of $P_i$ on the network relaying delay is defined as: when $P_i$ changes while the packet delivery probabilities of the other nodes keep constant, the variation of $DT_{G(1,2,\ldots n)}$, denoted as $\Delta DT^i_{G(1,2,\ldots n)}$.

According to the Definition 6 and (2), the $\Delta DT^i_{G(1,2,\ldots n)}$ (where $1 \leq i \leq n$ and $n$ is the degree of the relay network) can be calculated as:

$$\Delta DT^i_{G(1,2,\ldots n)} = \left(\sum_{i=1}^{n-1} iP_{i+1} \prod_{j=1}^{i}(1-(P_i+\Delta P))\right.$$
$$\left. + n \prod_{i=1}^{n}(1-(P_i+\Delta P))\right)T$$
$$= \begin{cases} (1-P_2) \cdot \left[\sum_{j=2}^{n-1}\left(j \cdot P_{j+1} \cdot \prod_{j=3}^{n-1}(1-P_j)\right)\right. \\ \qquad \left. + n \cdot \prod_{j=3}^{n}(1-P_j)\right]\Delta P \cdot T, \ i=1 \\ \left(\prod_{j=1}^{i-1}(1-P_j)\right) \cdot \left[\sum_{j=i}^{n-1}\left(j \cdot P_{j+1} \cdot \prod_{j=i+1}^{n-1}(1-P_j)\right)\right. \\ \qquad \left. + n \cdot \prod_{j=i+1}^{n}(1-P_j) - (i-1)\right]\Delta P \cdot T, 1 < i < n \end{cases} \quad (4)$$

where $P_j$ represents the packet delivery probability of the *jth* forwarding node in $G(1,2,\ldots n)$; $n$ is the degree of $G(1,2,\ldots n)$; $\Delta P$ is the variation of the packet delivery probability $P_i$. Note that the *j* used in (4) does not the node forwarding priority in $\mathbb{R}$, it is the forwarding priority in $\mathbb{R}^*$. For instance, if the relay network is $G(2,6,7)$, then the $P_1$, $P_2$, and $P_3$ in (4) represent $P_2$, $P_6$, and $P_7$, respectively. The coefficient of each term in (4) does not change for the same relay network. Based on (4), we can calculate the difference of the relaying delay variation between two adjacent forwarding nodes $\Delta DT^i_{G(1,2,\ldots n)}$ and $\Delta DT^{i+1}_{G(1,2,\ldots n)}$, which is denoted as $\Delta DT^{(i,i+1)}_{G(1,2,\ldots n)}$. The $\Delta DT^{(i,i+1)}_{G(1,2,\ldots n)}$ can be calculated as follows:

$$\Delta DT^{(i,i+1)}_{G(1,2,\ldots n)} = \Delta DT^i_{G(1,2,\ldots n)} - \Delta DT^{i+1}_{G(1,2,\ldots n)}$$
$$= \left(\prod_{j=1}^{i-1}(1-P_j)\right) \cdot \left[\cdot [1+(P_i-P_{i+1})\cdot[1+(1-P_{i+2})\right.$$
$$\left.\cdot\left(1+(1-P_{i+3})\cdots\left(1+(1-P_{n-1})(2-P_n)\right)\overset{n-i-2}{\cdots}\right)\right]\Delta P \cdot T$$
(5)

---
[2] The first-priority node is the node whose forwarding priority is 1; the second-priority node is the node whose forwarding priority is 2, and so on.

Based on (5), we can get the difference of the relaying delay variation between any two forwarding nodes, denoted as $\Delta DT^{(i,j)}_{G(1,2,\ldots n)}$, which is:

$$\Delta DT^{(i,j)}_{G(1,2,\ldots n)} = \sum_{k=i}^{j-1} \Delta DT^{(k,k+1)}_{G(1,2,\ldots n)}$$
$$= \left(\prod_{j=1}^{i-1}(1-P_j)\right) \cdot (1+(1-P_{i+1})$$
$$\cdot\left(1+(1-P_{i+2})\cdots\left(1+(P_i-P_j)(2-P_n)\right)\overset{j-i-2}{\cdots}\right)\Delta P \cdot T$$
(6)

For instance, for the relay network $G(1,2,3,7)$, $\Delta DT^{(1,3)}_{G(1,2,3,7)}$ represents the difference of the relaying delay variation between $\Delta DT^1_{G(1,2,3,7)}$ and $\Delta DT^3_{G(1,2,3,7)}$.

**Corollary 3:** To the relay networks which the priority of the forwarding nodes are determined based on the packet delivery probability based metrics, the higher forwarding priorities, the higher effect on the network relaying delay; i.e., if $i > j$, then $\Delta DT^{(i,j)}_{G(1,2,\ldots n)} > 0$ ; and if $(i-j) > (i-k)$ , then $\Delta DT^{(i,j)}_{G(1,2,\ldots n)} > \Delta DT^{(i,k)}_{G(1,2,\ldots n)}$.

**Proof.** This can be proved directly by (4), (5), and (6).

Corollary 3 demonstrates that the packet delivery probabilities of the higher-priority forwarding nodes have a greater effect on the network performance than that of the lower-priority forwarding nodes. Based on (4) and (5), we can derive the Corollary 4 and Corollary 5 as follows.

**Corollary 4:** To the relay networks which the forwarding priorities of the candidate forwarding nodes are decided based on the packet delivery probability based metrics, with the increasing of the network degree, the effect of the same $P_i$ becomes more and more serious; i.e., if $n > m$, then $\Delta DT^i_{G(1,2,\ldots n)} > \Delta DT^i_{G(1,2,\ldots m)}$ and $\Delta DT^{(i,j)}_{G(1,2,\ldots n)} > \Delta DT^{(i,j)}_{G(1,2,\ldots m)}$.

**Proof.** This can be proved directly by (4), (5), and (6).

For instance, based on Corollary 4, for the relay networks $G(1,2,3)$ and $G(1,2,3,7)$ , the $\Delta DT^{(1,3)}_{G(1,2,3)}$ is smaller than $\Delta DT^{(1,3)}_{G(1,2,3,7)}$ and the $\Delta DT^1_{G(1,2,3)}$ is smaller than $\Delta DT^1_{G(1,2,3,7)}$.

**Corollary 5:** To the relay network $G(1,2,\ldots n)$ which the priorities of the candidate forwarding nodes are decided based on the packet delivery probability based metrics, with the decreasing of the forwarding priority, if $n \to \infty$ , then $\Delta DT^{(i,i+1)}_{G(1,2,\ldots n)} \to 0$ and $\Delta DT^i_{G(1,2,\ldots n)} \to 0$.

**Proof.** See Appendix C.

The Corollary 5 demonstrates that the effect of the lower-priority forwarding node on the network performance becomes smaller and smaller when the number of nodes in the relay network increases.

For the relay networks which the node forwarding priorities are not decided based on the packet delivery probability relevant metrics, the properties are the same with that of the relay networks in which the node forwarding priorities are decided based on the packet delivery probability. Before investigating the properties of this kind of relay network, according to (5) and (6), we propose Corollary 6 first.

**Corollary 6:** To the relay network $G(1,2,\ldots,n)$ in which the forwarding priorities of the candidate forwarding nodes are not decided based on the packet delivery probability based metrics, if $P_i < P_j$, then the condition that $\Delta DT^{(i,j)}_{G(1,2,\ldots n)} < 0$ is true is shown as follows:

$$(P_j - P_i) > \frac{1+(2-P_{j-1})\cdot\prod_{k=i+1}^{j-2}(1-P_k)}{(2-P_{j+1})\cdot\prod_{k=i+1}^{j-1}(1-P_k)} = \varphi(i,j) > 1 \quad (7)$$

**Proof.** See Appendix D.

As shown in (7), since $P_i$ and $P_j$ are all smaller than 1, so the $(P_j - P_i)$ is smaller than 1, too. Therefore, (7) will not hold. The conclusion in Corollary 6 means that even $P_i < P_j$, then $\Delta DT^{(i,j)}_{G(1,2,\ldots n)} > 0$. Moreover, the Corollary 6 also illustrates that not only the packet delivery probability but also the forwarding priority can affect the network relaying delay.

Based on Corollary 6, we can conclude that to the relay networks in which the priorities of the forwarding nodes are decided based on the packet delivery probability irrelevant metrics, with the decreasing of the forwarding priority, the effect of the node packet delivery probability on the network relaying delay decreases. This means that for the network in which the nodes are prioritized based on the packet delivery probability irrelevant metrics, we can get the same corollaries as that shown in Corollary 3, Corollary 4, and Corollary 5.

According to the properties of the relay network, the parameters of the node whose forwarding priority is high have a greater effect on the transmission delay than that of the node whose priority is low. So, for reducing the transmission delay, the higher-priority forwarding nodes should have higher packet delivery probabilities than that of the lower-priority forwarding nodes. This conclusion is similar to the conclusions in [1] and [14]. In [14], the authors illustrate that the packet delivery probability of the node which is at the end of the communication link has a great effect on the energy consumption; the communication link which this packet delivery probability is low will deteriorate the routing performance greatly. The authors in [1] use the ETX which relates to all the packet delivery probabilities in the communication link to evaluate the effect on the routing performance. In this paper, we prove that the packet delivery probability of the higher-priority forwarding nodes can affect the transmission delay greatly.

Since for reducing the transmission delay, the higher-priority forwarding node should have a higher packet delivery probability than that of the lower-priority forwarding nodes. However, this is not always true in the algorithms which the node priority is not determined based on the packet delivery probability related metrics. In these algorithms, the high forwarding priority does not mean small packet delivery probability. For instance, if the performance metric is the residual energy, the node which has large residual energy may not have a higher packet delivery probability than the nodes whose residual energy is small. Therefore, for reducing the relaying delay, one approach is to re-set the forwarding priority based on the packet delivery probability. However, this will deteriorate the routing performance, because the node which the residual energy is large may be set a low forwarding priority if the packet delivery probability of this node is small. So, for taking both the node utility (which is used to determine the forwarding priority of node) and the packet delivery probability into account, the node priority needs to be re-calculated.

Assuming that the utility of the *ith*-priority forwarding node is $U_i$ ($U_i$ does not take the packet delivery probability into account) and the packet delivery probability of this node is $P_i$, we define the one-hop ETX for each forwarding nodes as follows: $ETX_{one-hop} = 1 / P_i$. Therefore, when taking the packet delivery probability into consideration, the utilities of the candidate forwarding nodes will deteriorate. The lower of the packet delivery probability, the more serious of the deterioration. Therefore, the new utility which has taken the packet delivery probability into account can be calculated as:

$$U_i^* = U_i / ETX_{one-hop} = U_i \cdot P_i \qquad (8)$$

The (8) demonstrates that when taking the packet delivery probability into account, the utility of forwarding node $i$ deduces to $U_i^*$ from $U_i$. The new priorities of the candidate forwarding nodes will be determined based on the value of $U_i^*$. An example can be found in Table 2. As shown in Table 2, when taking both the packet delivery probability and the residual energy into account, node *b* has better performance than node *a* and node *c*. In Table 2, we can find that the high-priority node determined by (8) has both high packet delivery probability and residual energy.

TABLE 2.
AN EXAMPLE

| node | 1 | 2 | 3 | 4 | 5 |
|---|---|---|---|---|---|
| residual energy (%) | 0.9 | 0.87 | 0.83 | 0.79 | 0.75 |
| packet delivery probability (%) | 0.65 | 0.78 | 0.8 | 0.69 | 0.57 |
| priority decided by residual energy | 1 | 2 | 3 | 4 | 5 |
| priority decided by packet delivery probability | 4 | 2 | 1 | 3 | 5 |
| priority decided by (8) | 3 | 1 | 2 | 4 | 5 |

V. NETWORK-BASED CANDIDATE FORWARDING SET OPTIMIZATION APPROACH

In Section III, we introduce the network model and the calculation model of the network relaying delay and packet delivery probability. In Section IV, we investigate the properties of the relay network, including the inter-network properties and in-network properties. In this section, based on the conclusions in Section III and Section IV, we propose the relay network recognition algorithm (RNR) and network-based candidate forwarding set optimization approach.

*A. Relay network recognition algorithm*

In Section III, we introduce the definition of the relay network, which is the fully connected sub-network of $G(V_\mathbb{R}, E_\mathbb{R})$. The relay networks include the *s*-networks and *o*-networks. Moreover, the *s*-networks can be derived from the *o*-networks. However, how to judge whether the nodes in $\mathbb{R}^*$ can construct a relay network or not has not been investigated sufficiently. In this section, based on the conclusion in Corollary 2, we propose a relay network recognition algorithm (RNR) to estimate whether any *n* nodes can constitute a relay network or not and distinguish the relay network is a *s*-network or *o*-network.

Before introducing RNR, we first define the neighbor matrix for each candidate forwarding node. Assuming that there are *m* nodes in $\mathbb{R}$, for node *i*, the neighbor matrix can be expressed as:

$$\begin{array}{c} 1\,2\,3\,4\cdots i\cdots m \\ neib_i = [0\,1\,0\,0\cdots 1\cdots 1] \end{array} \qquad (9)$$

In (9), if the node *j* has a bi-directional communication link with node *i*, then the *jth* value in $neib_i$ will be "1"; otherwise, this value will be "0". In RNR, we regard that node *i* is a neighbor of itself. For estimating the existence of the relay network, we define a sum operator between any two neighbor matrixes as follows.

**Definition 7:** For two neighbor matrixes which only contain "0" and "1", the "+" between two neighbor matrixes $neib_i$ and $neib_j$ is defined as:

$$D_{(i,j)} = neib_i + neib_j = \sum_{k=1}^{m} \left( neib_i(k) \wedge neib_j(k) \right) \quad (10)$$

where "∧" is the "and" operator in Boolean algebra. For instance, to the matrixes [1 0 0 1 1 1] and [0 1 0 1 1 0], based on (10), the summary of these two matrixes will be 2. According to Definition 7, we can estimate whether any *n*-degree network is the relay network or not.

**Corollary 7:** For any network $G(V_\mathbb{R}, E_\mathbb{R})$ which the network degree is *n*, if $D_{G(V_R, E_\mathbb{R})} \geq n$, then the network is the relay network; otherwise, the network is not the relay network.

**Proof.** See Appendix E.

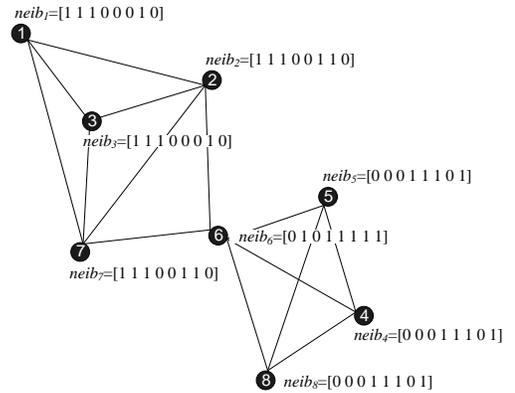

Fig. 4. The neighbor matrixes of the candidate forwarding nodes in Fig. 2(a)

For instance, the neighbor matrixes of the candidate forwarding nodes in Fig. 2(a) are shown in Fig. 4. As shown in Fig. 4, according to the Definition 7, $D_{G(1,2,3)} = 4$, which is larger than its network degree, so based on the Corollary 7, we can conclude that $G(1,2,3)$ is a relay network. However, since $D_{G(2,5,6)} = 1$, which is smaller than its network degree, so $G(2,5,6)$ is not a relay network. The rest of the relay networks can be gotten by the same process based on the conclusions of Definition 7 and Corollary 7. Note that the relay networks gotten from Corollary 7 include both the *s*-networks and *o*-networks. The Corollary 7 can only be used to estimate whether the network is the relay network or not; it cannot distinguish the *s*-network from the *o*-network. Therefore, we propose the Corollary 8 to distinguish different kinds of relay networks.

**Corollary 8:** For any relay network $G(V_{\mathbb{R}^*})$ which the network degree is $n$, if $D_{G(V_{\mathbb{R}^*})} = n$, then the network $G(V_{\mathbb{R}^*})$ is an $o$-network; otherwise, if $D_{G(V_{\mathbb{R}^*})} = m > n$, where $n$ is the degree of $G(V_{\mathbb{R}^*})$, then $G(V_{\mathbb{R}^*})$ is a $s$-network, and the degree of the $o$-network that $G(V_{\mathbb{R}^*})$ is derived from is $m$; moreover, based on (3), the number of the relevant $m$-degree $s$-network is $c_m^n$.

**Proof.** See Appendix E.

For instance, in Fig. 4, $D_{G(1,2,3)} = 4$ and the degree of $G(1,2,3)$ is 3, so $G(1,2,3)$ is $s$-network and derived from an $o$-network which the network degree is 4. Additionally, the number of 3-degree relevant $s$-network of $G(1,2,3)$ is $c_4^3 = 4$. In Fig. 4, since $D_{G(1,2,3,7)} = 4$ which is equal to its network degree, so the network $G(1,2,3,7)$ is an $o$-network.

The relay network recognition algorithm is shown as follows.

**Algorithm 1:** The Relay Network Recognition (RNR) Algorithm
1. candidate forwarding node $i$ calculates the neighbor matrix $neib_i$;
2. if $D_{G(V_{\mathbb{R}^*})_n} = n \rightarrow G(V_{\mathbb{R}^*})_n$ is the $o$-network;
3. if $D_{G(V_{\mathbb{R}^*})_n} = m > n \rightarrow G(V_{\mathbb{R}^*})_n$ is the $s$-network;
4. if $D_{G(V_{\mathbb{R}^*})_n} < n \rightarrow G(V_{\mathbb{R}^*})_n$ is not the relay network.

*B. Network parameters calculation*

After the recognition of the relay networks, we need to decide which relay network is the most appropriate one as the final relay network. The nodes in the selected relay network will be the final forwarding nodes and the other nodes in $\mathbb{R}$ will be deleted.

As talked in Section I, for improving the performance of opportunistic routing, during the relay network selection, the following properties of the relay network should be met as much as possible: 1) the forwarding delay of the relay network should be as small as possible; 2) the packet delivery probability of the relay network should be as large as possible; 3) the network in which the utilities of forwarding nodes (i.e. the forwarding priorities of forwarding nodes) are high should be selected as much as possible to guarantee high network performance. Therefore, in the relay network selection, not only the network packet delivery probability and the network forwarding delay, but also the node utilities in the relay network should be considered.

Based on (1) and (2), the forwarding delay and packet delivery probability of forwarding network can be calculated, respectively. According to the Expect Transmission Count (ETX) defined in [1] and the packet delivery probability of network, we define the one-hop ETX of the forwarding network $G(1,2,...,n)$ as:

$$t_{G(1,2,...n)} = \frac{1}{P_{G(1,2,...n)}} = \frac{1}{1-\prod_{i=1}^{n}(1-P_i)} \quad (11)$$

where $P_i$ is the packet delivery probability of node $i$ in the forwarding network. When takes the one-hop network ETX into account, the network forwarding delay deteriorates, which can be calculated as:

$$DT^*_{G(1,2,...,n)} = DT_{G(1,2,...n)} \cdot t_{G(1,2,...n)}$$
$$= \frac{\left(\sum_{i=1}^{n-1} iP_{i+1}\prod_{j=1}^{i}(1-P_j)+n\prod_{i=1}^{n}(1-P_i)\right)T}{1-\prod_{i=1}^{n}(1-P_i)} \quad (12)$$

Similar to the analysis in Section III, during the relay network selection, the relay network which has good performances on both network forwarding delay and node utilities should have a high priority to be selected as the final relay network. For evaluating the effect of node utilities on network performances, we define and calculate the network utility $U_{G(1,2,...n)}$ as follows.

For the relaying network $G(1,2,...,n)$, considering the packet delivery probabilities and utilities of forwarding nodes in the relay network, the network utility $U_{G(1,2,...n)}$ varies. This can be expressed in (13):

$$U_{G(1,2,...,n)} = \begin{cases} U_1, & \text{the probability is } P_1 \\ U_2, & \text{the probability is } P_2(1-P_1) \\ \vdots \\ U_n, & \text{the probability is } P_n \prod_{i=1}^{n-1}(1-P_i) \\ 0, & \text{the probability is } \prod_{i=1}^{n}(1-P_i) \end{cases} \quad (13)$$

where $U_i$ is the utility of the *ith* forwarding node. Therefore, for the relay network whose network degree is $n$, the average network utility can be calculated as:

$$\bar{U}_{G(1,2,...,n)} = U_1 \cdot P_1 + \sum_{i=2}^{n}\left(U_i \cdot \prod_{j=1}^{i-1}(1-P_j)P_i\right) \quad (14)$$

The (14) is the average network utility of network $G(1,2,...,n)$ on one transmission try. Similar to the network relaying delay, when taking the network ETX which calculated in (11) into account, this utility deteriorates. The network utility which takes the one-hop network ETX into account can be calculated as:

$$U^*_{G(1,2,...,n)} = \frac{\bar{U}_{G(1,2,...,n)}}{t_{G(1,2,...,n)}}$$
$$= \left(U_1 \cdot P_1 + \sum_{i=2}^{n}\left(U_i \cdot \prod_{j=1}^{i-1}(1-P_j)P_i\right)\right)$$
$$\cdot \left(1 - \prod_{i=1}^{n}(1-P_i)\right) \quad (15)$$

Based on (12) and (15), we can find that for each relay network, two network parameters should be taken into account during the relay network selection: the network forwarding delay $DT^*_{G(1,2,...,n)}$ and the network utility $U^*_{G(1,2,...,n)}$. Both these two parameters take the network ETX into account, so the packet delivery probability of the relay network is also considered indeed.

*C. Best relay network selection*

The selected relay network should have high quality performances on both of the two parameters calculated in Section V.B. In this paper, for achieving this purpose, we introduce the weight based multi-attribute utility theory into the final network utility calculation. The multi-attribute utility theory is effective in dealing with this kind of issue [3][18]. Based on this approach, the final network utility can be calculated as:

$$U^F_{G(1,2,...,n)} = \omega_{DT} \cdot DT^*_{G(1,2,...,n)} + \omega_U \cdot U^*_{G(1,2,...,n)} \quad (16)$$

where $\omega_{DT}$ is the weight of $DT^*_{G(1,2,...,n)}$ and $\omega_U$ is the weight of $U^*_{G(1,2,...,n)}$.

For the weight based multi-attribute utility theory, the first important issue is to determine the weights of each performance metrics. To the metrics of the relay network, there is a fact that the metric (i.e., $DT^*_{G(1,2,...,n)}$ and $U^*_{G(1,2,...,n)}$) whose variance is large has a greater effect on the network performance than that of the small one [3][4][18]. For instance, as the parameters which are shown in Table 3, since the values of $U^*_{G(1,2,...,n)}$ between different relay networks are similar, which $U^*_{G(1,2,...,n)}$ is chosen has a small effect on the network performances. However, for different relay networks, the values of $DT^*_{G(1,2,...,n)}$ are quite different, so which $DT^*_{G(1,2,...,n)}$ is chosen has a great effect on the network performances. Based on this conclusion, one of the feasible approaches is to use the variances of $DT^*_{G(1,2,...,n)}$ and $U^*_{G(1,2,...,n)}$ as the weights in (16).

However, as shown in [3], [4], and [18], if we use the values of $DT^*_{G(1,2,...,n)}$ and $U^*_{G(1,2,...,n)}$ that calculated in (12) and (15), and the variances of $DT^*_{G(1,2,...,n)}$ and $U^*_{G(1,2,...,n)}$ directly, there are problems. Because: 1) the final network utility will be decided mainly by the parameter whose value is large; for instance, in Table 3, since the value of $DT^*_{G(1,2,...,n)}$ is much larger than that of $U^*_{G(1,2,...,n)}$, the value of $U^F_{G(1,2,...,n)}$ will be decided mainly by $DT^*_{G(1,2,...,n)}$; 2) the variance is affected seriously by the value of the parameter, so it cannot reflect the practical variation rate of this parameter; for instance, as shown in Table 3, the variance of $U^*_{G(1,2,...,n)}$ is larger than that of $DT^*_{G(1,2,...,n)}$; however, when taking the values of the parameters into account, the variation rate of $U^*_{G(1,2,...,n)}$ is smaller than that of $DT^*_{G(1,2,...,n)}$ in fact. So, when we choose the next hop relay network, the $DT^*_{G(1,2,...,n)}$ has a greater effect on the routing performances than that of the $U^*_{G(1,2,...,n)}$. This is because the variance is the absolute difference between different parameters, so it is affected seriously by the values of parameters. Therefore, in this paper, for investigating the effect of different parameters on the routing performances, we propose the concept of relative variance (*rv*) and use the relative variance as the weight of the parameter [3][18].

TABLE 3.
AN EXAMPLE

| network | a | b | c | variance | rv |
|---|---|---|---|---|---|
| $U^*_{G(1,2,...,n)}$ | 51 | 52 | 53 | 0.67 | 0.00074 |
| $DT^*_{G(1,2,...,n)}$ | 0.27 | 0.68 | 0.49 | 0.028 | 0.366 |

The relative variance is defined as:
$$v_{r_x} = \frac{1}{k}\sum_{i=1}^{k}\left(\frac{x_i - \bar{x}}{\bar{x}}\right)^2 \quad (17)$$
where $x$ represents $DT^*_{G(1,2,...,n)}$ or $U^*_{G(1,2,...,n)}$, $\bar{x}$ is the average of $x$, $k$ is the number of relay networks. In the relative variance, the value of (17) can reflect the effect of different parameters on the routing performances accurately. This can be found in Table 3. In Table 3, even the variance of $U^*_{G(1,2,...,n)}$ is larger than that of $DT^*_{G(1,2,...,n)}$, the relative variance of $DT^*_{G(1,2,...,n)}$ is larger than that of $U^*_{G(1,2,...,n)}$, which is consist with the effect of the parameter on the routing performances.

For evaluating the difference between the relative variances of these two metrics, we define the parameter resolution ratio $\xi$ as:
$$\xi = \begin{cases} \frac{v_{rDT}}{v_{rU}}, & v_{rDT} > v_{rU} \\ 1, & v_{rDT} = v_{rU} \\ \frac{v_{rU}}{v_{rDT}}, & v_{rDT} < v_{rU} \end{cases} \quad (18)$$

From (18), we can find that $\xi \geq 1$, the larger $\xi$, the larger difference between the relative variances of these two parameters. For the network utility calculated in (16), with the increasing of $\xi$, the effect of the metric whose relative variance is large on the network utility increases; the effect of the parameter whose variance is small on the network utility decreases. When the $\xi$ is small, the effect of these two parameters on the network utility is similar.

For the first issue, if we use the values of the parameters directly in the network utility calculation, there are problems. For instance, as the metrics which are shown in Table 3, since the relative variance of Metric_1 is smaller than that of the Metric_2, according to the analysis above, the network utility should be affected mainly by the Metric_2. However, the fact is that the network utilities are decided mainly by Metric_1, i.e., the network in which the value of Metric_1 is the largest will have the highest network utility. According to the network utility defined in (16), the priorities of the network utilities are: *network_c→network_b→network_a*, which is the same as the priorities of Metric_1. This is not consistent with the analysis above. The reason is that the value of Metric_1 is much larger than that of the Metric_2. When the difference between Metric_1 and Metric_2 is too large, it will cover up the effect of Metric_2 on the relay network selection. For solving this issue, in [4] and [18], the authors map the different order of magnitudes parameters to the same order of magnitude. In this paper, considering the fact that for each performance metric, there is an order number that relates to them, we introduce the order number of the parameter into the network utility calculation. For instance, based on the values of Metric_2 shown in Table 4, the order numbers of the Metric_2 are 1, 3, and 2 for network *a*, *b*, and *c*, respectively. The large order number means that the related metric's value is large in the relay network, vice versa. So, in this paper, the value of the parameter shown in (16) will be replaced by the order number of the parameter, which can be expressed as:
$$U^F_{G(1,2,...,n)} = v_{rDT} \cdot n^i_{DT_{G(1,2,...,n)}} + v_{rU} \cdot n^i_{U_{G(1,2,...,n)}} \quad (19)$$
where $n^i_{DT_{G(1,2,...,n)}}$ is the order number of $DT$ in $G(1,2,...,n)$, $n^i_{U_{G(1,2,...,n)}}$ is the order number of $U$ in $G(1,2,...,n)$. The final network utility will be decided by (19), which can be found in Table 4. In Table 4, the network utility of network_b is larger than that of network_c, which is consistent with the analysis above. In Table 4, we also present the network utilities that calculated based on the algorithm proposed in [18] (which is the weight based algorithm) and [19] (which is the fuzzy logic based algorithm). From Table 4, we can find that the priorities of the relay networks calculated by (19) are the same as those calculated by [18] and [19].

TABLE 4.
AN EXAMPLE

| network | a | b | c | rv |
|---|---|---|---|---|
| Metric_1 | 29 | 45 | 63 | 0.0925 |
| Order number of Metric_1 | 1 | 2 | 3 | |
| Metric_2 | 0.27 | 0.68 | 0.49 | 0.122 |
| Order number of Metric_2 | 1 | 3 | 2 | |
| Utility calculated by (16) | 2.72 | 4.25 | 5.89 | |
| Utility calculated by (19) | 0.3365 | 0.551 | 0.3995 | |
| Utility calculated by [18] | 0.06 | 0.125 | 0.118 | |
| Utility calculated by [19] | 0.448 | 0.529 | 0.517 | |

Based on (18) and (19), we can derive the property of this algorithm as follows. The network utility calculated by (19) relates to both the weight and the order number of the metric. Assuming that there are two relay networks, for the network_a, the order number based on $DT$ is $e_i$ and the order number based on $U$ is $e_j$; for the network_b, the order numbers relate to these two metrics are $e_m$ and $e_k$, respectively. Let $\Delta^e_{DT} = |e_i - e_j|$, $\Delta^e_U = |e_m - e_k|$, $\xi = \alpha$, and $v_{rDT} > v_{rU}$, then we can derive the property of this algorithm as follows.

**Corollary 9.** If $\Delta^e_U/\Delta^e_{DT} < \alpha$, the utility will be decided mainly by $DT$; if $\Delta^e_U/\Delta^e_{DT} > \alpha$, the utility will be decided mainly by $U$; vice versa.

**Proof.** See Appendix F.

An example can be found in Fig. 6. The values of the metrics in Fig. 6 are the same as that shown in Table 3. As shown in Fig. 6(a), for the network_b and network_c, since $\Delta_{DT} = 1$ and $\Delta_U = 1$, so $\Delta_U/\Delta_{DT} = 1$; since $\xi = 1.32 > \Delta_U/\Delta_{DT}$, so in Fig. 6(a), the network utility will be decided mainly by the value of $DT$. Therefore, in Fig. 6(a), the forwarding priority of network_b is 1 and the priority of network_c is 2, which is the same as the order of $DT$. However, as shown in Fig. 6(b), for network_b and network_c, since $\Delta_U/\Delta_{DT} = 2$ which is larger than $\xi$, so the network utility will be decided mainly by the value of $U$. Therefore, in Fig. 6(b), the forwarding priorities of network_b and network_c are 2 and 1, respectively; this is the same as the order of $U$. The forwarding priorities of network_b and network_c are opposite in these two figures.

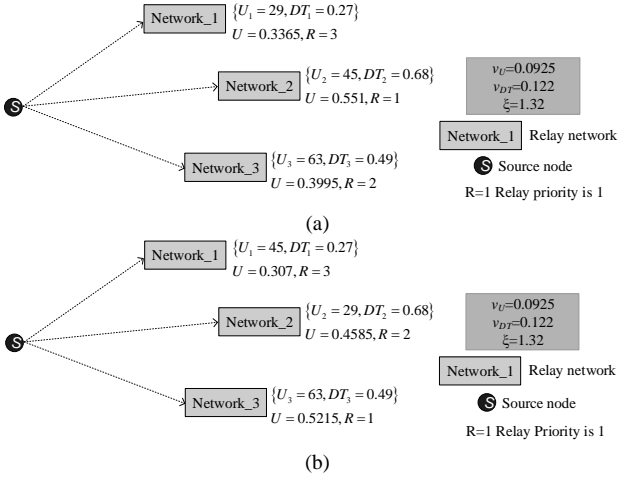

Fig. 6. An example of the relay network priority and selection algorithm

When the relay network is selected by the algorithm introduced above, the nodes in the relay network will forward the packet based on the forwarding priority calculated in Section IV. The coordination scheme is time-based, which has been introduced in [1], [5], and [8] in detailed. The waiting timer is set to 45*ms*, which is the same as that shown in [8]. When the node in the relay network forwards the data packet to the next hop relay network, the processes are the same as that shown above. This process will be repeated until the data packet is received by the destination node. The process of the network-based CFS optimization algorithm can be found below.

---

**Algorithm 2:** DDA based CFS optimization algorithm

1. each relay network calculates the one-hop network ETX based on (11);
2. based on (12) and (15), the network relaying delay $DT^*_{G(1,2,\ldots,n)}$ and network utility $U^*_{G(1,2,\ldots,n)}$ are calculated;
3. the source node calculates the variances of network relaying delay and network utility, i.e., $v_{DT^*_{G(1,2,\ldots,n)}}$ and $v_{U^*_{G(1,2,\ldots,n)}}$, respectively;
4. applying the Corollary 2, Corollary 3, and Corollary 4 to pre-selecting the relay network;
5. based on (19), the final network utility $U^F_{G(1,2,\ldots,n)}$ is calculated;
6. the relay network which has the highest final network utility will be chosen as the final relay network.

---

Moreover, we also investigate the computational complexity of the proposed network-based candidate forwarding set optimization algorithm.

**Corollary 10.** The computational complexity of the algorithm proposed in this paper is $kO(n^2)$, where $k$ is the number of relay networks, and $n$ is the number of nodes in the relay network.

**Proof.** See Appendix G.

## VI. SIMULATION AND DISCUSSION

In this section, we will evaluate the performances of DDA. We compare the performances of DDA with ExOR [1] and SOAR [8], respectively. The variables are the number of nodes and the number of CBR connections. The number of CBR connections represents the traffic load of the network. The parameters of the simulation environment are shown in Table 5.

TABLE 5.
SIMULATION PARAMETERS

| simulation parameter | value |
|---|---|
| simulation area | 2000*m*×2000*m* |
| number of nodes | 100, 150,…, 300 |
| maximum transmission range | 250*m* |
| channel data rate | 1Mbps |
| the traffic type | Constant Bit Rate (CBR) |
| number of CBR connections | 20, 40,…, 100 |
| packet size | 512bytes |
| beacon interval | 1*s* |
| maximum packet queue length | 50 packets |
| MAC layer | IEEE 802.ll DCF |
| TTL | 32 |
| simulation tool | NS2 |

The performance matrixes used in this paper are the transmission delay, the packet delivery ratio, and the network throughput: (1) *end-to-end packet delivery ratio*: the packet delivery ratio is defined as the ratio of the number of packets received successfully by the destination node to the number of packets generated by the source node [14][20]; (2) *end-to-end delay*: the transmission delay of the data packet from the source node to the destination node; (3) *network throughput*: the network throughput is the ratio of the total number of packets received successfully by the destination node to the number of packets sent by all the nodes during the simulation time [21].

### A. Performance under different network densities

In this section, we evaluate the performance of DDA, SOAR, and ExOR under different network densities, i.e., the number of nodes in the network varies. In this simulation, the network load is constant, so the number of the CBR connections is set to 60. The results can be found in Fig. 7, Fig. 8, and Fig. 9.

In Fig. 7, the average end-to-end delay of these three algorithms is presented. In these three algorithms, with the increase of the number of nodes, the average end-to-end delay decreases. The fewer nodes in the network, the larger the decrease. For instance, in DDA, when the number of nodes increases from 100 to 150, the delay decreases from 780*ms* to 602*ms*. However, when the number of nodes increases from 250 to 300, the delay decreases from 520*ms* to 500*ms*. Similar conclusions can be found in SOAR and ExOR. This can be explained as: when the number of nodes increases, the probability of network portion decreases, so the delay will decrease; when the network density is large enough, the probability of network portion is quite low, so the decreasing of the transmission delay becomes slowly. Moreover, for the same network density, the end-to-end delay of DDA is much smaller than that of the other two algorithms. For instance, when the number of nodes is 100, the delay of DDA is 26.6% smaller than that of SOAR and 35.4% smaller than that of ExOR, respectively. When the number of nodes is 200, the delay of DDA is 14.5% smaller than that of SOAR and 45.5% smaller than that of ExOR, respectively. This is because in DDA, the relay nodes are fully connected and the relay network which the delay is small has a high priority to be chosen, so the end-to-end delay in DDA is the smallest in these three algorithms.

In Fig. 8, the packet delivery ratios of these three algorithms are illustrated. With the increase of the network density, the packet delivery ratios of these three algorithms increase. Because with the increasing of the network density, for the sender, more and more forwarding nodes can be found in its transmission range; so according to (1), the network packet delivery ratio increases. The packet delivery ratio of DDA is the largest in these three algorithms. For instance, when the number of nodes is 150, the packet delivery ratio of DDA is 19% larger than that of SOAR and 23.8% larger than that of ExOR, respectively. When the number of nodes is 300, the packet delivery ratio of DDA is 9% larger than that of SOAR and 21.3% larger than that of ExOR, respectively. Since in DDA, the packet delivery ratio is taken into account during the relay network selection, so the packet delivery ratio of DDA is the largest. In Fig. 8, when the network density is large enough, this increasing becomes slowly. For instance, when the number of nodes increases from 100 to 200, the packet delivery ratios of DDA and SOAR increase 41.8% and 28.8%, respectively; however, when the number of nodes increases from 200 to 300, the packet delivery ratios of DDA and SOAR increase 14.1% and 20.9%, respectively. This is because when the network density is large enough, the number of forwarding nodes is large; so there always exits at least one node can receive the data packet and send it to the destination node. This makes increasing slow.

The network throughput of these three algorithms is presented in Fig. 9. From Fig. 9, we can conclude that when the network density increases, the network throughput keeps constant approximately. These values fluctuate in a very small range. For instance, approximately, the variation range is 0.03 in DDA and 0.02 in SOAR. On one hand, when the network density is small, the packet delivery ratio is small (which can be found in Fig. 8), so the probability of retransmission is high. However, the number of hops to the destination is small when the network density is small, which contributes to the reduction of the number of control packets. On the other hand, when the network density is large, the packet delivery ratio increases; however, the average number of hops to the destination node increases. This causes the increasing of the number of control packets. So, the network throughput keeps stable in these algorithms. Moreover, the network throughput of DDA is the best in these three algorithms. For instance, when the number of nodes is 100, the network throughput of DDA is 16.2% larger than that of SOAR and 32.4% larger than ExOR, respectively. When the number of nodes is 300, the network throughput of DDA is 19.5% larger than that of SOAR and 36.5% larger than ExOR, respectively.

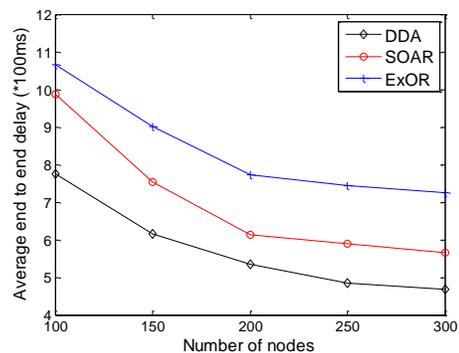

Fig. 7. The average end to end delay under different network densities.

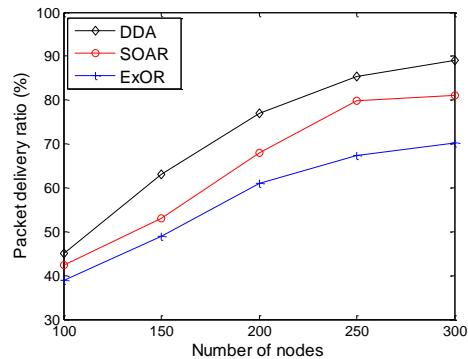

Fig. 8. The packet delivery ratio under different network densities.

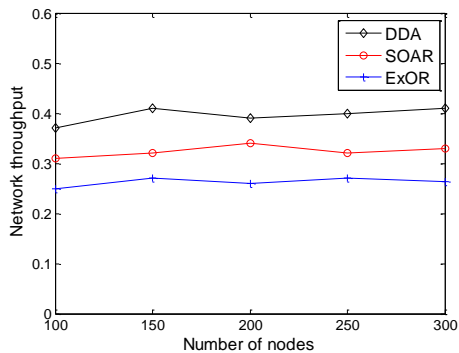

Fig. 9. The network throughput under different network densities.

*B. Performance under different traffic load*

In this section, the performances of these three algorithms under the different number of CBR connections are presented. In this simulation, the number of nodes in the network is 200 and the number of CBR connections varies. The results can be found in Fig. 10, Fig. 11, and Fig. 12.

In Fig. 10, the average end-to-end delay of these three algorithms is shown. The results of the end-to-end delay under different traffic loads are different from those under different network densities. With the increasing of the number of CBR connections, the end-to-end delay is the smallest when the number of CBR connections is 60. The delay decreases when the number of CBR connections smaller than 60 and increases when the number of CBR connections larger than 60. For instance, when the number of CBR connections increases from 20 to 60, the end-to-end delay of DDA and SOAR decreases by 24.6% and 35.1%, respectively; however, when the number of CBR connections increases from 60 to 100, the end-to-end delay of DDA and SOAR increases by 20.3% and 28.4%, respectively. This is because with the increasing of the traffic load, when the number of CBR connections is not large enough

(for instance, smaller than 100), the network resources are far from saturated. So, when the traffic load increases, the end-to-end delay decreases. However, when the number of CBR connections is large enough, the network becomes saturated or over-saturated, so the network competition becomes more and more serious. This will deteriorate the performances of the algorithms. Moreover, the end-to-end delay of DDA is the smallest in these three algorithms. For instance, when the number of CBR connections is 60, the end-to-end delay of DDA is 12.7% smaller than that of SOAR and 30.4% smaller than ExOR, respectively. When the number of CBR connections is 100, the end-to-end delay of DDA is 10.2% smaller than that of SOAR and 33% smaller than ExOR, respectively.

When the traffic load increases, the packet delivery ratios of these three algorithms decrease, which can be found in Fig. 11. The reason is that when the number of CBR connections increases, network competition becomes more and more serious. Moreover, similar to Fig. 10, when the number of CBR connections is small, this decreasing is slow. However, when the number of CBR connections is large, this decreasing is fast. For instance, when the number of CBR connections increases from 20 to 40, the packet delivery ratios of DDA and SOAR decrease by 2.4% and 1.4%, respectively; however, when the number of CBR connections increases from 80 to 100, the packet delivery ratios of DDA and SOAR decrease by 10.6% and 15.9%, respectively. This is because when the number of CBR connections is small, the network resources, such as the buffer of each node, are not saturated; so even the network competition and the network interference increase, the decreasing of the packet delivery ratio is slow. However, when the network resource is saturated or over-saturated, the network interference and the network competition increase, so the decreasing of the packet delivery ratio becomes more and more serious. Moreover, the packet delivery ratio of DDA is the best in these three algorithms. For instance, when the number of CBR connections is 20, the packet delivery ratio of DDA is 15.3% larger than that of SOAR and 18.8% larger than ExOR, respectively. When the number of CBR connections is 60, the packet delivery ratio of DDA is 11% larger than that of SOAR and 15.1% larger than ExOR, respectively.

The network throughput of these three algorithms under a different number of CBR connections is shown in Fig. 12. Different from that shown in Fig. 9, the network throughput which is shown in Fig. 12 decreases when the network traffic load increases. The decreasing of the network throughput can be explained by Fig. 10 and Fig. 11. When the number of CBR connections increases, on one hand, the end-to-end delay decreases at first and increases after the inflection point (i.e. 100); on the other hand, when the traffic load increases, the packet delivery ratio decreases. Additionally, when the traffic load increases, the network interference, the network competition, and the channel occupation ratio increase seriously, so the network throughput decreases. However, as shown in Fig. 12, the decreasing of ExOR and SOAR is much faster than that of DDA. Moreover, the network throughput of DDA is the largest in these three algorithms. For instance, when the number of CBR connections increases from 20 to 100, the network throughput of DDA, SOAR, and ExOR decrease 11.9%, 17.1%, and 37.1%, respectively. This is because the duplicated transmission in the time-based coordination scheme is reduced as much as possible in DDA, which contributes to the increase of the network throughput. Moreover, the network throughput of DDA is the best in these three algorithms. For instance, when the number of CBR connections is 20, the network throughput of DDA is 2.4% larger than that of SOAR and 16.7% larger than that of ExOR. When the number of CBR connections is 100, the network throughput of DDA is 8.1% larger than that of SOAR and 40.5% larger than that of ExOR.

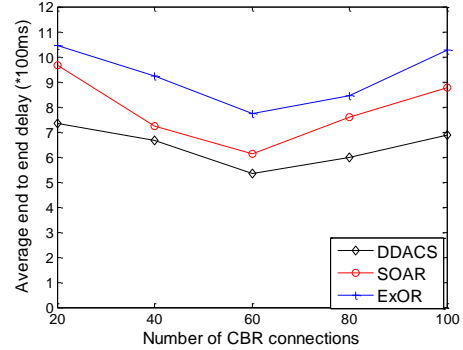

Fig.10. The average end-to-end delay under different traffic loads.

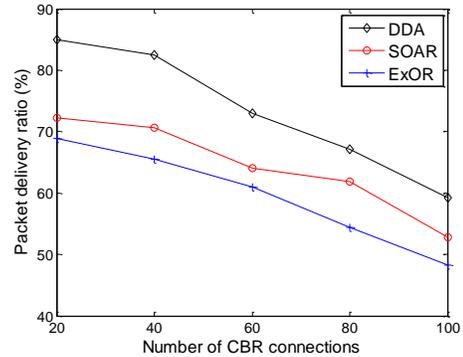

Fig.11. The packet delivery ratio under different traffic loads.

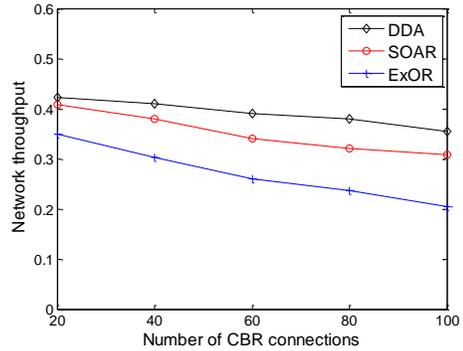

Fig.12. The network throughput under different traffic loads.

## VII. CONCLUSION

In this paper, for improving the performances of opportunistic routing, we propose the network-based candidate forwarding set optimization algorithm for the time-based coordination scheme. In this algorithm, the candidate forwarding nodes are divided into different fully connected relay networks, so the duplicated transmission is avoided. Moreover, in this paper, we also propose the RNR algorithm which can be used to judge whether the sub-network is fully connected or not. When the fully connected relay networks are gotten, then these relay networks will be used as the basic units in the next hop relay network selection. In this paper, we prove that the packet delivery probability of the high-priority forwarding nodes in the relay network has a greater effect on the relaying delay than that of the low-priority forwarding nodes. According to this conclusion, in DDA, the relay network has high forwarding priority, if the packet delivery probabilities of the high-priority forwarding nodes are high in this relay network. During the next hop relay network selection, the transmission delay, the network utility, and the packet delivery probability are taken into consideration. By these innovations, the DDA can improve the network performance greatly than ExOR and SOAR. Moreover, in this paper, the properties of the relay networks are investigated in detail.